# Interlayer sliding direction as a symmetry selector in altermagnetic bilayer $Fe_2WS_4$: Switchable anomalous Hall and anomalous valley Hall effects

Quan Shen, Jianing Tan, Tao Yao, Wenhu Liao, and Jiansheng Dong*

*Department of Physics, Jishou University, Jishou 416000, Hunan, China*

*Corresponding author: jsdong@jsu.edu.cn

**ABSTRACT**

Altermagnets combine compensated collinear magnetic order with momentum-dependent spin splitting, offering a promising platform for coupling spin and valley degrees of freedom with ferroelectricity and Berry-curvature driven transport in the absence of net magnetization. However, achieving nonvolatile and selective control of these intertwined degrees of freedom remains a key challenge. Here, using first-principles calculations, we show that the direction of interlayer sliding serves as a symmetry selective control parameter in altermagnetic bilayer $Fe_2WS_4$. Diagonal sliding breaks inversion symmetry and produces two sliding ferroelectric states with opposite out-of-plane polarizations. Reversal of the ferroelectric polarization switches the momentum-dependent spin texture and reverses the anomalous Hall conductivity, revealing strong magnetoelectric coupling and enabling a ferroelectrically switchable anomalous Hall effect. In contrast, axial sliding preserves inversion symmetry but breaks the crystalline symmetry relating the X and Y valleys, leading to reversible valley polarization and a switchable anomalous valley Hall effect. These results establish the direction of interlayer sliding as a nonvolatile symmetry selector for controlling ferroelectricity, spin texture, valley polarization, and Hall transport responses in two-dimensional altermagnetic bilayers.

## I. INTRODUCTION

Altermagnets (AMs) have recently emerged as a distinct class of compensated collinear magnets that combine zero net magnetization with momentum-dependent spin splitting [1,2]. Unlike conventional antiferromagnets (AFMs), in which opposite-spin sublattices are connected by spatial inversion ( $P$ ) or lattice translation symmetry and thus exhibit spin-degenerate bands, AMs host nonrelativistic spin splitting because their opposite-spin sublattices are instead related by crystal rotations or mirror symmetries [3-6]. This symmetry mechanism combines the robustness of compensated magnetic order with spin-polarized electronic states, making AMs a promising platform for AFM spintronics and Berry-curvature driven transport phenomena [7-9]. Importantly, because their spin splitting does not rely primarily on spin-orbit coupling (SOC), AMs offer an appealing route to robust spin-polarized responses in compensated magnetic systems.

Beyond spin, the valley degree of freedom provides an additional information channel in two-dimensional materials [10-12]. In conventional valleytronic materials, valley degeneracy is generally protected by time-reversal ($T$) symmetry, giving rise to $T$-paired valleys; lifting this degeneracy therefore typically requires $T$ symmetry breaking. In AMs, by contrast, valley degeneracy can be protected by crystalline symmetry, giving rise to crystalline-symmetry-paired ($C$-paired) valleys [13]. This distinct symmetry origin makes valleys in AMs particularly sensitive to symmetry-lowering perturbations. To date, valley polarization in AMs has mainly been induced by uniaxial strain [14,15], which breaks the crystalline symmetry connecting the $C$-paired valleys [16]. However, strain-based control requires mechanical deformation and directional strain manipulation, which are not ideal for reversible, nonvolatile, and electrically addressable device operation. A symmetry selective control strategy that avoids external mechanical deformation is therefore highly desirable for practical valleytronic applications in AMs.

Sliding ferroelectricity (FE) in van der Waals bilayers provides a natural route to such symmetry selective control [17-20]. Interlayer translation changes the lattice registry and can lower selected crystal symmetries without chemical modification, enabling nonvolatile control of electronic states while preserving structural integrity. Recent studies have shown that sliding FE can switch spin polarization and anomalous Hall responses [21-23]. These results suggest that interlayer sliding is not merely a structural displacement, but can act as an internal symmetry breaking perturbation in two-dimensional materials. However, whether the sliding direction itself can serve as a symmetry selector for valley polarization remains unresolved. More generally, it remains unclear whether interlayer sliding can selectively couple FE to spin, valley, and Hall transport responses within a single AM platform.

In this work, using first-principles calculations, we show that AM bilayer $Fe_2WS_4$ realizes sliding direction selective coupling. Monolayer $Fe_2WS_4$ hosts sizable nonrelativistic spin splitting and intrinsic spin-valley locking. In the AA-stacked bilayer, $P$ symmetry, interlayer mirror ($M_z$) symmetry, and diagonal mirror ($M_{xy}$) symmetry collectively protect spin degeneracy and valley equivalence. Diagonal sliding breaks $P$ symmetry and produces two sliding FE states with opposite out-of-plane polarizations; reversal of this polarization switches the momentum-dependent spin texture and reverses the anomalous Hall conductivity, realizing a switchable anomalous Hall effect (AHE). In contrast, axial sliding preserves $P$ symmetry while breaking the crystalline symmetry relating the X and Y valleys, yielding reversible valley polarization and a switchable anomalous valley Hall effect (AVHE). These results establish interlayer sliding as a nonvolatile, symmetry-selective control knob for manipulating FE, spin texture, valley polarization, and Hall transport responses in AM bilayers.

## II. CALCULATION DETAILS

First-principles calculations were performed within the framework of density functional theory (DFT) using the Vienna *ab initio* Simulation Package (VASP) [24,25]. The projector-augmented-wave (PAW) method was employed to describe the electron-ion interaction, and the Perdew-Burke-Ernzerhof (PBE) parametrization of the generalized gradient approximation (GGA) was adopted for the exchange-correlation functional [26]. Structural optimizations of $Fe_2WS_4$ systems were carried out with convergence criteria of $10^{-6}$ eV for the total energy and 0.001 eV Å$^{-1}$ for the atomic forces. A plane-wave cutoff energy of 500 eV was adopted, and a 30 Å vacuum layer was introduced along the out-of-plane direction to eliminate spurious interactions between periodic images. To account for van der Waals interactions in the bilayer system, Grimme's DFT-D3 correction was applied [27]. Brillouin-zone integrations were performed using a Γ-centered Monkhorst-Pack *k*-point grid of $9\times9\times1$ for both structural relaxation and electronic structure calculations. To account for the correlation effects of Fe 3*d* electrons, the DFT+*U* method was employed with an effective Hubbard parameter $U_{eff}=2.0$ eV [28,29], where $U_{eff}=U-J$ and $J=0$. Thermal stability was assessed through *ab initio* molecular dynamics (AIMD) simulations in a $5\times5\times1$ supercell at 300 K for 5 ps. Dynamical stability was confirmed by phonon-dispersion calculations using the PHONOPY code [30,31]. Berry-curvature distributions were calculated using the VASPBERRY code [32]. To evaluate the anomalous Hall conductivity, maximally localized Wannier functions were constructed using the WANNIER90 package [33-35], with Fe and W *d* orbitals and S *p* orbitals as initial projections, and the anomalous Hall conductivity was obtained from Wannier interpolation.

## III. RESULTS AND DISCUSSION

### A. Monolayer $Fe_2WS_4$: Structural Stability and Electronic Structure

Monolayer $Fe_2WS_4$ crystallizes in a tetragonal structure with space group $p\bar{4}2m$ (No. 111) and optimized in-plane lattice constants of $a$ = $b$ = 5.47 Å. As shown in Figs. S1(a) and S1(b), the top view reveals a square lattice preserving the $M_{xy}$ symmetry, whereas the side view shows a sandwich-like S-(Fe, W)-S atomic framework. The small energy fluctuations observed during the AIMD simulation at 300 K [Fig. S1(c)], together with the absence of appreciable imaginary frequencies throughout the Brillouin zone [Fig. S1(d)], confirm the thermal and dynamical stability of monolayer $Fe_2WS_4$.

To determine the magnetic ground state, we considered three representative collinear magnetic configurations, namely FM, AFM1, and AFM2, using a $\sqrt{2}\times\sqrt{2}\times1$ supercell [Figs. S2(a)-S2(c)]. Following previous studies, an effective Hubbard parameter $U_{eff}=2.0\ \text{eV}$ was applied to the Fe 3$d$ orbitals within the DFT+$U$ framework [36,37]. Total energy calculations identify AFM1 as the ground state. In this configuration, the Fe magnetic moments are aligned antiparallel, resulting in zero net magnetization. The AFM coupling is mainly attributed to S-mediated super-exchange between neighboring Fe atoms [36,38].

Although the two Fe sublattices are magnetically compensated in real space, they are related by the $M_{xy}$ symmetry, rather than by $P$ or a lattice translation. This symmetry relation allows momentum-dependent spin splitting even in the absence of SOC. As shown in Fig. S1(e), the spin-resolved band structure exhibits pronounced nonrelativistic spin splitting, with the X and Y valleys carrying opposite spin characters and thereby giving rise to intrinsic spin-valley locking. Including SOC does not qualitatively alter the band structure [Fig. S1(f)], confirming that the spin splitting is primarily nonrelativistic in origin. These results establish monolayer $Fe_2WS_4$ as a stable AM parent platform for investigating sliding controlled valley polarization and Hall responses.

**B. Bilayer $Fe_2WS_4$: Stacking and Sliding Configurations**

Stacking two-dimensional magnetic materials into van der Waals bilayers introduces an additional interlayer degree of freedom, enabling electronic, magnetic, and FE phenomena beyond those of isolated monolayers, including stacking-induced AM [39,40], interlayer magnetoelectric coupling [41,42], sliding FE [18-20], and tunable Hall transport [43,44]. Owing to the weak interlayer interaction, lateral sliding between adjacent layers provides an energy efficient, controllable, and nonvolatile route to manipulate these coupled physical properties.

We first constructed the high-symmetry AA-stacked bilayer $Fe_2WS_4$ as the reference structure. As shown in Fig. S3, total energy calculations indicate that interlayer AFM coupling is energetically favored. In the AA stacking, the $M_z$ symmetry exchanges the two antiferromagnetically coupled layers without changing the in-plane momentum $k$. Therefore, it connects opposite-spin states at the same momentum, $M_z E(s,k) = E(-s,k)$, thereby enforcing spin degeneracy at each $k$. In addition, $P$ symmetry relates electronic states at opposite momenta, $PE(s,k) = E(s,-k)$, thereby ensuring their energetic equivalence [45,46]. The coexistence of these symmetries thus imposes the combined $M_z P$ band constraint, $M_z PE(s,k) = E(-s,-k)$, which enforces degeneracy between opposite-spin states, while the preserved $M_{xy}$ symmetry interchanges the X and Y valleys and thereby ensures their valley degeneracy. Collectively, these symmetries render the AA-stacked bilayer nonpolar, spin-degenerate, and valley-equivalent, consistent with the calculated spin-resolved band structure [Fig. S4(b)].

Starting from the AA-stacked bilayer, we fixed the bottom layer and laterally translated the top layer to generate four representative sliding configurations. The translation vectors are expressed in fractional coordinates with respect to the in-plane lattice vectors. The $AB_1$ and $AB_2$ configurations are obtained through diagonal translations $\tau_1 = (\frac{1}{4}, \frac{1}{4}, 0)$ and $\tau_2 = (-\frac{1}{4}, \frac{1}{4}, 0)$, respectively. In contrast, $AC_1$ and $AC_2$ are generated by axial translations $\tau_3 = (\frac{1}{2}, 0, 0)$ and $\tau_4 = (0, \frac{1}{2}, 0)$, corresponding to sliding along the *x* and *y* directions, respectively. The top views and sliding paths of these configurations are shown in Fig. S5, whereas schematic illustrations of the $AB_1$/$AB_2$ and $AC_1$/$AC_2$ configurations are presented in Figs. 1(a) and 1(d), respectively. After full structural relaxation, the four sliding configurations exhibit comparable energies and retain their characteristic stacking geometries [Fig. S6], indicating that they are metastable states accessible via interlayer sliding. This metastability provides the structural basis for the subsequent analysis of their electronic, FE, and Hall responses.

As schematically summarized in Fig. 1, these metastable sliding configurations provide two distinct control channels. Along the diagonal sliding path, $AB_1$ and $AB_2$ form a pair of polar states with opposite out-of-plane FE polarizations [Fig. 1(a)]. Switching between these two states reverses the momentum-dependent spin polarization and the anomalous Hall response [Figs. 1(b) and 1(c)]. By contrast, axial sliding generates the $AC_1$ and $AC_2$ configurations [Fig. 1(d)]. These two states are related by a symmetry operation that exchanges the *x* and *y* directions and exhibits opposite X/Y valley polarizations [Fig. 1(e)]. The corresponding valley-contrasting Berry-curvature response gives rise to a switchable AVHE [Fig. 1(f)]. Overall, these results establish interlayer sliding as a symmetry selective control knob for switching FE, spin, valley, and Hall responses in bilayer $Fe_2WS_4$.

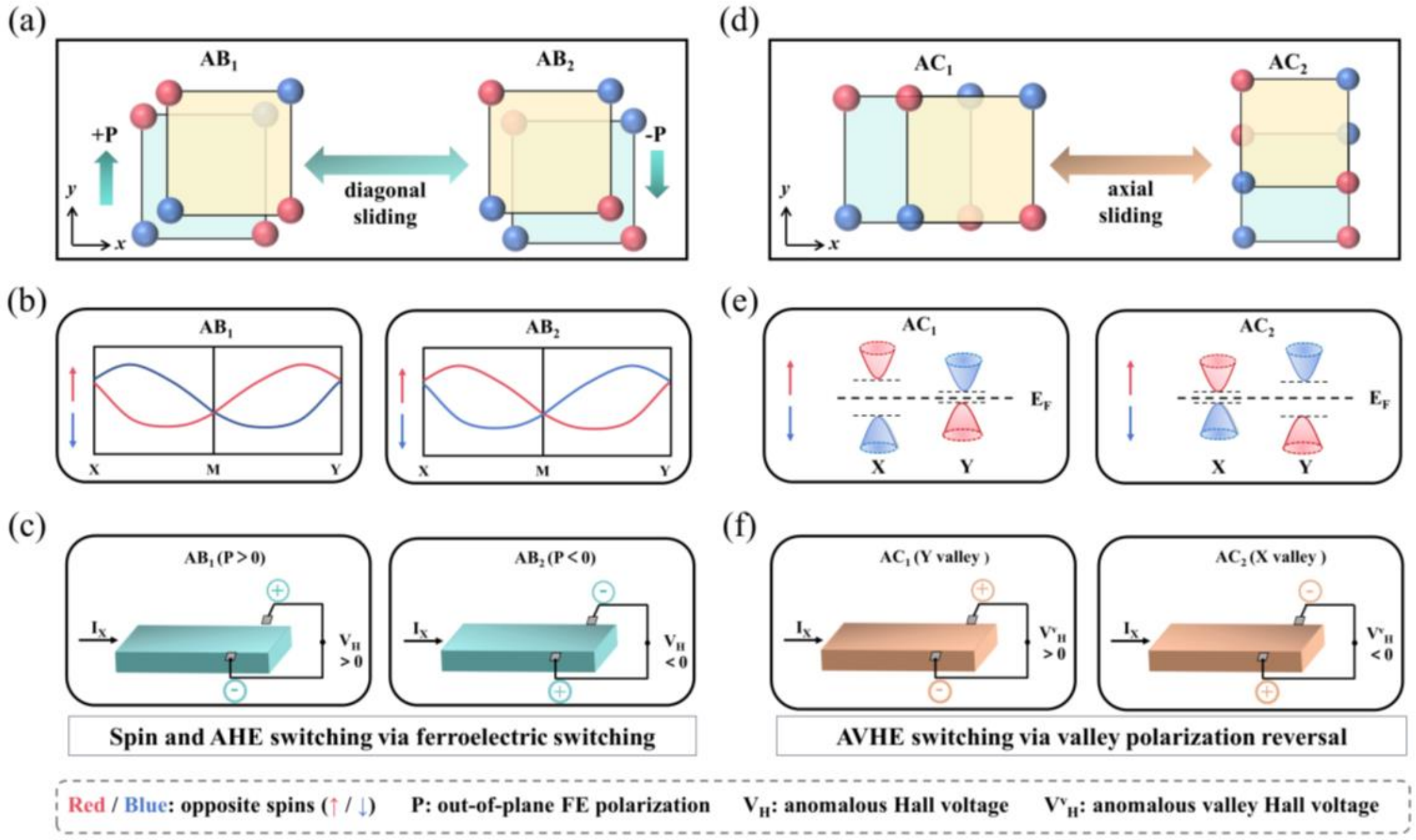

**FIG. 1.** Schematic illustration of sliding direction selective coupling in bilayer $Fe_2WS_4$. (a) Diagonal sliding generates the polar $AB_1$ and $AB_2$ configurations with opposite out-of-plane FE polarizations. (b) FE polarization reversal between $AB_1$ and $AB_2$ switches the momentum-dependent spin polarization, revealing magnetoelectric coupling. (c) The concomitant reversal of the anomalous Hall response realizes a switchable AHE. (d) Axial sliding produces the $AC_1$ and $AC_2$ configurations. (e) Switching between $AC_1$ and $AC_2$ reverses the valley polarization between the X and Y valleys. (f) This valley polarization reversal gives rise to a switchable AVHE.

## C. Diagonal Sliding: Ferroelectricity and AHE Switching

We first examine the $AB_1$ and $AB_2$ configurations generated by diagonal interlayer sliding from the AA stacking. Unlike the centrosymmetric AA stacking, diagonal interlayer sliding breaks $P$ symmetry, thereby allowing a finite out-of-plane FE polarization $P_z$ in the AB configurations [18-20]. This polar character is further confirmed by the plane-averaged electrostatic potential and charge redistribution shown in Fig. S7. For $AB_1$, the electrostatic potential exhibits a finite vacuum-level offset of $\Delta\Phi = 15\ \mathrm{meV}$ [Fig. S7(a)], accompanied by an asymmetric interlayer charge redistribution [Fig. S7(b)]. In contrast, $AB_2$ shows an opposite potential step, $\Delta\Phi = -15\ \mathrm{meV}$ [Fig. S7(c)], together with a reversed charge-transfer profile [Fig. S7(d)]. These results demonstrate that $AB_1$ and $AB_2$ form a pair of sliding FE states with equal-magnitude but oppositely directed out-of-plane polarizations.

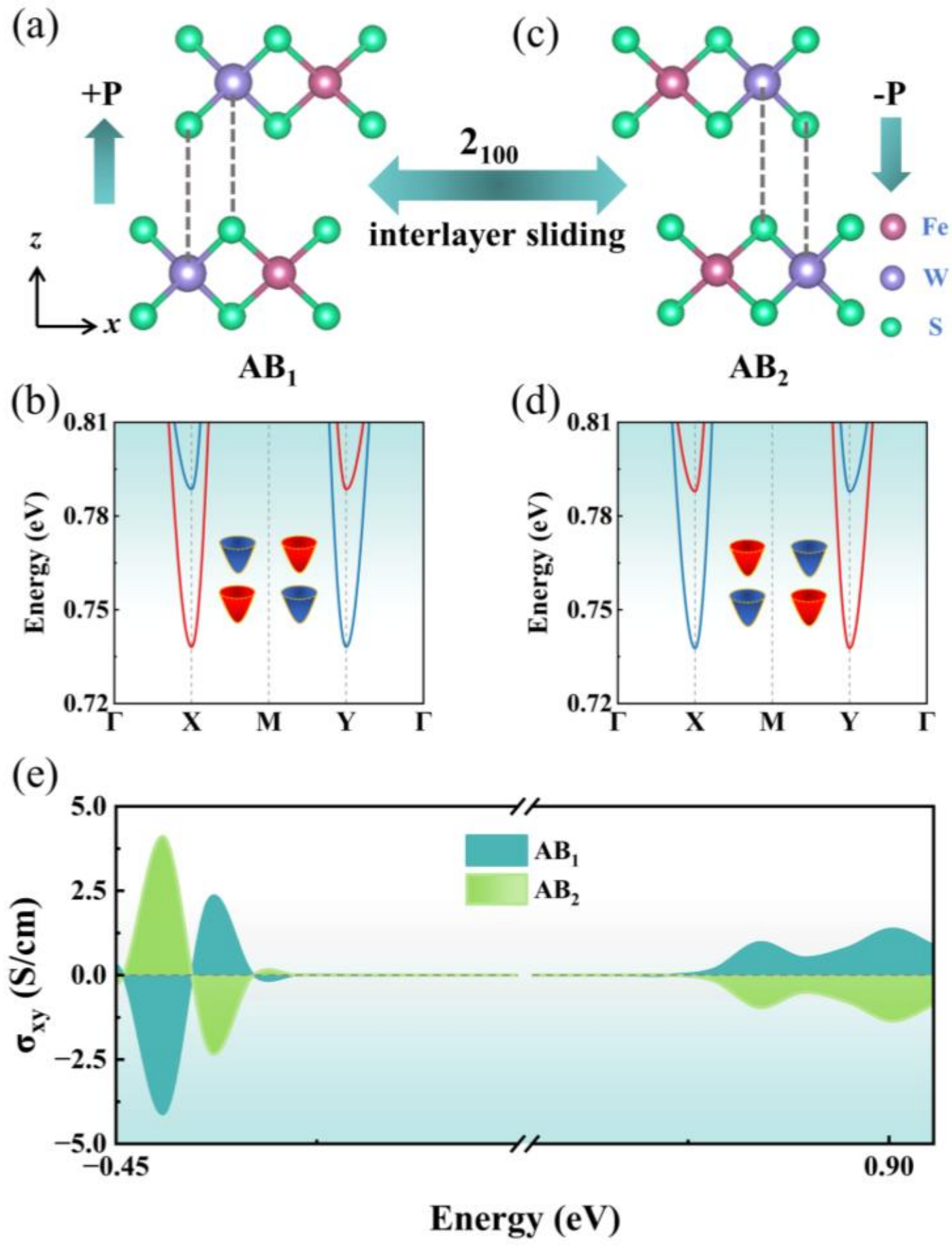

**FIG. 2.** (a),(c) Side views of the $AB_1$ and $AB_2$ configurations, which are connected by the $2_{100}$ operation and exhibit opposite out-of-plane FE polarizations. (b),(d) Spin-resolved band structures of $AB_1$ and $AB_2$ without SOC, showing reversed spin characters between the two polar states. Red and blue denote opposite spin channels. (e) Intrinsic anomalous Hall conductivity $\sigma_{xy}$ of $AB_1$ and $AB_2$ calculated with SOC as a function of energy, showing opposite signs upon $AB_1/AB_2$ switching.

The relation between the two polar states is further governed by symmetry. As shown in Figs. 2(a) and 2(c), $AB_1$ and $AB_2$ are connected by the $2_{100}$ operation. In real space, $2_{100}$ corresponds to a twofold rotation about the *x* axis, transforming (*x, y, z*) into (*x, -y, -z*). Because the *z* coordinate is reversed, this operation exchanges the upper and lower layers and reverses the sign of any out-of-plane polar vector. For the relative interlayer displacement, the layer exchange introduces an additional sign change, so the displacement pattern of $AB_1$ is mapped onto that of $AB_2$ under $2_{100}$. Consequently, $2_{100}$ connects the two sliding-related FE states and reverses their out-of-plane polarizations, giving $P_z(AB_1) = -P_z(AB_2)$. Thus, $AB_1$ and $AB_2$ constitute a pair of symmetry related sliding FE states.

We next examine how this FE switching affects the AM spin texture. In the absence of SOC, the spin and lattice degrees of freedom are decoupled. In this limit, $AB_1$ and $AB_2$ are related by the spin-space connection operation $N_S = |2\perp||2_{100}|$, where $2\perp$ denotes a twofold spin rotation perpendicular to the collinear magnetic moments [47]. This operation gives $E_{AB_1}(s,k) = E_{AB_2}(-s, 2_{100}k)$. Since $2_{100}$ transforms the in-plane momentum as $(k_x, k_y) \rightarrow (k_x, -k_y)$, the relation becomes $E_{AB_1}(s, k_x, k_y) = E_{AB_2}(-s, k_x, -k_y)$. This symmetry relation explains the calculated band structures in Figs. 2(b) and 2(d): $AB_1$ and $AB_2$ have nearly identical band dispersions, whereas their spin characters are reversed. Therefore, switching the FE polarization between $AB_1$ and $AB_2$ simultaneously switches the momentum-dependent AM spin polarization, revealing strong magnetoelectric coupling.

When SOC is included, spin is coupled to the lattice and an independent spin rotation is no longer allowed. The connection between $AB_1$ and $AB_2$ must then be described within the magnetic space group. For the magnetic configuration considered here, the $2_{100}$ operation itself reverses the spin direction and connects $AB_1$ with $AB_2$; thus the relevant magnetic connection operation becomes $N_m = 2_{100}$. Under this operation, the out-of-plane Berry curvature satisfies $\Omega_z^{AB_1} = -\Omega_z^{AB_2}$. The intrinsic anomalous Hall conductivity is obtained by integrating the Berry curvature over the occupied states in the Brillouin zone, $\sigma_{xy} = -\frac{e^2}{\hbar}\sum_n \int_{BZ} \frac{d^2k}{(2\pi)^2} f(\varepsilon_{nk})\Omega_{n,z}(k)$ [48], where $f(\varepsilon_{nk})$ is the occupation function. The above Berry-curvature relation therefore requires the anomalous Hall conductivity to change sign, $\sigma_{xy}^{AB_1} = -\sigma_{xy}^{AB_2}$. This symmetry requirement is consistent with Fig. 2(e), where $AB_1$ and $AB_2$ exhibit anomalous Hall conductivities with opposite signs. Thus, diagonal sliding enables the simultaneous nonvolatile switching of FE polarization, momentum-dependent spin polarization, and the AHE in bilayer $Fe_2WS_4$.

**D. Axial Sliding: Valley Polarization and AVHE Switching**

We next examine the axial sliding path that generates the $AC_1$ and $AC_2$ configurations. As shown in Figs. 3(a) and 3(b), these two stackings are connected by an in-plane translation of the top layer relative to the bottom layer. In contrast to the diagonal AB stackings, axial sliding preserves *P* symmetry while breaking the $M_{xy}$ symmetry. For $AC_1$ with the relative displacement $\tau = (\frac{1}{2}, 0, 0)$, inversion transforms τ into $-\tau = (-\frac{1}{2}, 0, 0)$, which is translationally equivalent to $\tau$ after adding an in-plane lattice vector. The same argument applies to $AC_2$ with $\tau = (0, \frac{1}{2}, 0)$, for which $-\tau = (0, -\frac{1}{2}, 0)$ is also equivalent to the original displacement. Therefore, both $AC_1$ and $AC_2$ retain *P* symmetry, which forbids a net out-of-plane FE polarization. This conclusion is further supported by the plane-averaged electrostatic potentials in Fig. S8, where no vacuum-level offset is observed for either configuration.

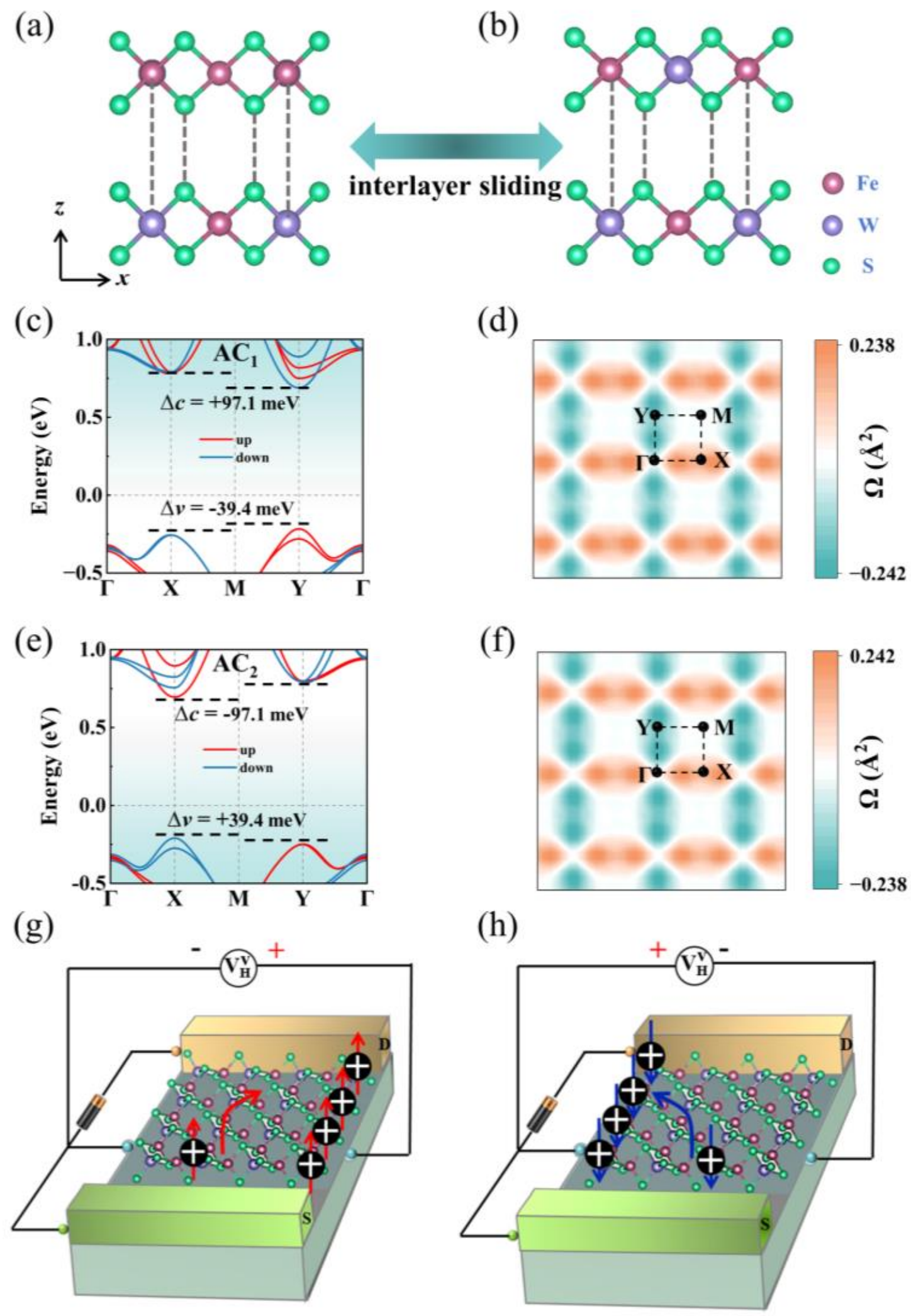

**FIG. 3.** (a),(b) Side views of the $AC_1$ and $AC_2$ configurations connected by axial interlayer sliding and the $M_{xy}$ operation. (c),(e) Spin-resolved band structures of $AC_1$ and $AC_2$ without SOC, showing reversed valley polarization. Red and blue denote opposite spin channels. (d),(f) Berry curvature $\Omega_z$ of the valence band for $AC_1$ and $AC_2$ calculated with SOC. (g),(h) Schematic illustration of the switchable anomalous valley Hall response in $AC_1$ and $AC_2$.

Although the axial AC stackings are nonpolar, each individual configuration breaks the $M_{xy}$ symmetry connecting the X and Y valleys, thereby lifting their degeneracy and inducing finite valley polarization. We define the conduction- and valence-band valley polarizations as $\Delta_{C(V)} = E_{XC(V)} - E_{YC(V)}$, where $E_{XC(V)}$ and $E_{YC(V)}$ denote the conduction-band minimum (CBM) or valence-band maximum (VBM) energies at the X and Y valleys, respectively. As shown in Fig. 3(c), $AC_1$ exhibits $\Delta_C = +97.1$ meV and $\Delta_V = -39.4$ meV. Switching to $AC_2$ reverses the signs of the valley polarizations, yielding $\Delta_C = -97.1$ meV and $\Delta_V = +39.4$ meV [Fig. 3(e)]. This reversal follows directly from the $M_{xy}$ symmetry connection between $AC_1$ and $AC_2$, which exchanges the X and Y valleys. Therefore, axial interlayer sliding provides a nonvolatile route to reverse valley polarization without generating out-of-plane FE.

We further analyze the valley dependent Berry curvature, which governs anomalous transverse carrier motion in two-dimensional systems. The Berry curvature is calculated from the Kubo formula [48]:

$$\Omega_z(k) = -\sum_n \sum_{n \neq m} f_n(\vec{k}) \frac{2\,\mathrm{Im}\langle \psi_{nk} | \hat{\upsilon}_x | \psi_{mk} \rangle \langle \psi_{mk} | \hat{\upsilon}_y | \psi_{nk} \rangle}{\left(E_{nk} - E_{mk}\right)^2} \tag{1}$$

where $f_n(k)$ is the Fermi-Dirac distribution function, $\psi_{nk}$ ($\psi_{mk}$) is Bloch wave function with eigenvalue $E_{nk}$ ($E_{mk}$), and $\hat{\upsilon}_{x/y}$ is the velocity operator along the *x*/*y* directions, respectively. As shown in Figs. 3(d) and 3(f), the valence band Berry curvature is concentrated around the X and Y valleys and exhibits a pronounced valley contrasting distribution. For $AC_1$, the Berry curvature reaches $\Omega_Z(X) = +0.238$ Å$^2$ and $\Omega_Z(Y) = -0.242$ Å$^2$ [Fig. 3(d)]. In $AC_2$, the corresponding values become $\Omega_Z(X) = +0.242$ Å$^2$ and $\Omega_Z(Y) = -0.238$ Å$^2$. Thus, the Berry curvatures at the two valleys retain opposite signs, while their relative magnitudes are interchanged by axial sliding, consistent with the reversal of valley polarization.

Under an in-plane electric field, the Berry curvature generates an anomalous carrier velocity, $v \approx -\frac{q}{\hbar} E \times \Omega(k)$ [49], where $E$ is the applied electric field. Because the X and Y valleys carry opposite Berry curvatures with unequal magnitudes, a finite Berry-curvature imbalance emerges near the valence band edge. This imbalance drives valley dependent anomalous transverse carrier motion and gives rise to a switchable AVHE, as schematically illustrated in Figs. 3(g) and 3(h). These results demonstrate that axial sliding selectively controls valley polarization and the AVHE, providing a distinct control channel from diagonal sliding, which switches FE, spin texture, and the AHE.

## IV. CONCLUSION

In summary, using first-principles calculations, we have demonstrated that interlayer sliding in an AM bilayer $Fe_2WS_4$ is not merely a structural displacement but a direction-dependent symmetry control mechanism that activates distinct coupled responses. Starting from the high-symmetry AA-stacked bilayer, where the combined $M_zP$ symmetry protects spin degeneracy and $M_{xy}$ preserves valley equivalence, different sliding directions select distinct symmetry breaking channels. For diagonal sliding, $P$ symmetry is broken, generating two oppositely polarized states. FE reversal between these states switches the momentum-dependent spin texture and reverses the anomalous Hall conductivity, thereby revealing magnetoelectric coupling and realizing a FE switchable AHE. For axial sliding, $P$ symmetry is retained, whereas the crystalline symmetry relating the X and Y valleys is broken, leading to reversible valley polarization and a switchable AVHE. These findings identify direction selective interlayer sliding as a structural symmetry selector for separately addressing FE polarization, spin texture, valley polarization, and Hall transport responses within a single AM bilayer. More broadly, this mechanism provides a design principle for nonvolatile, Hall-readable multifunctional states in two-dimensional magnetic van der Waals materials.

**ACKNOWLEDGMENTS**

This work was supported by the National Natural Science Foundation of China (Grant Nos. 12364007, 12264016).

**SUPPLEMENTAL MATERIAL**

See the Supplemental Material for details on monolayer stability and electronic structure, magnetic configurations, AA-bilayer symmetry, sliding pathways, relative stacking energies, and electrostatic-potential analyses of AB and AC stackings.

**DATA AVAILABILITY**

The data that support the findings of this article are available from the corresponding author upon reasonable request.